\documentclass[a4paper,12pt]{article}

\newcommand{\D}{\mathrm{d}}
\newcommand{\E}{\mathrm{e}}
\newcommand{\J}{\mathrm{i}}
\newcommand{\lfunc}{\mathcal{L}}
\newcommand{\rfunc}{\mathcal{R}}
\newcommand{\efunc}{\mathcal{E}}
\newcommand{\rbar}{\overline{r}}
\newcommand{\tbar}{\overline{\theta}}

\usepackage{amsmath}
\usepackage{amssymb}
\usepackage{times}
\usepackage{graphicx}
\usepackage{url}
\usepackage{cite}
\usepackage[squaren]{SIunits}

\graphicspath{{./Figures/}}

\title{Shielding of rotor noise by plates and wings}

\author{Michael Carley}

\begin{document}

\title{Shielding of rotor noise by plates and wings}

\author{Michael Carley\\Department of Mechanical
  Engineering\\University of Bath\\Bath BA2 7AY\\United
  Kingdom\\m.j.carley@bath.ac.uk}

\maketitle

\begin{abstract}
  A method of noise reduction proposed for the next generation of
  aircraft is to shield noise from the propulsion system, by
  positioning the noise source over a wing or another surface. In this
  paper, an approximate analysis is developed for the acoustic field
  far from a circular source placed near the edge of a semi-infinite
  plate, a model problem for shielding of noise by a wing and for
  scattering by a trailing edge. The approximation is developed for a
  source of small radius and is found to be accurate when compared to
  full numerical evaluation of the field.
\end{abstract}

\section{Introduction}
\label{sec:intro}

The effect of airframe configurations on the propagation of noise from
propulsion systems is a question which has been studied over a number
of years. In particular, studies have been conducted of the
possibility of positioning propulsion systems in such a way that the
airframe acts to ``shield'' the noise source and reduce noise levels
reaching the ground. One example is the ``Silent Aircraft''
Initiative, in which a study was conducted of the scattering of noise
from sources placed above the blended wing body
configuration~\cite{agarwal-dowling07}. Another
study~\cite{eret-kennedy-amoroso-castellini-bennett16} examined the
effect of propeller positioning on an otherwise conventional airframe,
making use of the tailplane for acoustic shielding. As electric
propulsion improves, and novel configurations using airframe shielding
for noise reduction become more feasible, there is a greater need for
reliable methods which can be used in design and to provide insight
into the propagation of sound around airframes.

The literature on the use of aircraft configurations to shield noise
from propulsion systems is not as extensive as that on the reduction
of noise at source, and has tended to concentrate on reducing
environmental noise from jet and turbofan systems. Czech and
Thomas~\cite{czech-thomas13} note that there have been very few
studies on the installation effect on open rotors of airframe elements
other than pylons. They report experimental studies of noise from an
open rotor system on a conventional and on a hybrid wing body
airframe, as part of an extensive study on installation and shielding
effects~\cite{bahr-thomas-lopes-burley-van-zante14,guo-czech-thomas13,%
  guo-thomas16}. The airframe configuration used in these tests had a
swept trailing edge and so was not comparable to the model problem
considered in this paper, but it did demonstrate shielding effects of
up to~12\deci\bel\ for certain axial displacements of the
rotor~\cite{czech-thomas13}. 

The prediction of shielding effects on propulsion noise appears to
have been carried out mainly using numerical approaches such as the
use of computational fluid dynamics (CFD) data in a permeable surface
integration method~\cite{duerrwaechter-kessler-kraemer19} or a
Boundary Element Method (BEM) calculation using a suitable model for
the rotor source~\cite{lummer-richter-proeber-delfs13}. A comparison
of shielding patterns from BEM predictions and experimental data has
also shown good agreement in the case of a monopole point
source~\cite{lummer-moessner-delfs18} and gives confidence that the
BEM is a suitable method for assessment of shielding when sufficient
information about the aircraft configuration is available. 

There does not appear to be a reliable method in the literature for
approximate prediction of shielding effects. In a study of shielding
of noise from a counter-rotation open rotor~\cite{stephens-envia11},
two theories for shielding attenuation, barrier theory and half-plane
diffraction, were compared to experimental data, approximating the
rotor by a point source. It was found that neither gave accurate
results, over-predicting the attenuation by~5\deci\bel\ when the point
source approximation was used. The error may have been increased by
the source having a higher frequency than was used in developing the
curve fits used in the barrier theory approximations~\cite{li-wong05}.

The theory for shielding by a wing is based on the problem of
diffraction by a semi-infinite plane, first addressed more than a
century ago~\cite{macdonald15}. A comprehensive review of the theory
of noise shielding~\cite{jones77} discusses in some depth the analysis
of noise from point sources near shielding bodies and the different
approximations which apply throughout the field. In particular, the
applicability and breakdown of the geometric acoustics approximation
is considered, as well as the existence of ``shadow'' zones around a
shielding body, regions in which the source is not ``visible'', being
blocked by the body. In this paper, the theory to be applied will be
the standard analysis of scattering by a semi-infinite plate, extended
to include the effects of a mean
flow~\cite{roger-moreau-kucukcoskun16}.


The motivation for the work in this paper is then the lack of a
reliable model for shielding of noise from rotating sources in
aeronautical applications. We present an analysis of far-field noise
from a rotating source which replaces the point source approximation
with the models normally used in established theory for noise from
rotating sources. The result is an approximation which compares well
to full numerical solution and should be useful to designers and
researchers as a starting point for parametric studies of rotor
positioning in new configurations.

\section{Analysis}
\label{sec:analysis}

The problem of scattering by a plane is an old
one~\cite{macdonald15,jones77}, which has been extended to include the
effect of mean flow in the acoustic
case~\cite{roger-moreau-kucukcoskun16}. We begin by presenting
existing results for the problem in a form which are then incorporated
into standard methods for the calculation of the acoustic far field of
rotating sources.

\subsection{Green's function}
\label{sec:analysis:greens}

\begin{figure}[t]
  \includegraphics{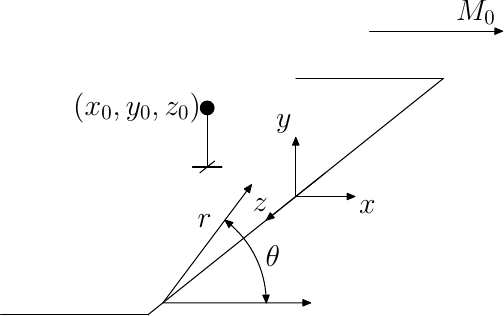}
  \caption{Notation for semi-infinite plate scattering}
  \label{fig:analysis:notation}
\end{figure}

The coordinate system used to analyze the scattering of sound by a
semi-infinite plate is shown in Figure~\ref{fig:analysis:notation}. A
uniform flow of Mach number $M_{0}$ is imposed normal to the trailing
edge. A Cartesian coordinate system $(x,y,z)$ is centred on the edge
of the plate, with $x$ parallel to the flow, $y$ normal to the plate
and $z$ aligned with the edge. A cylindrical coordinate system
$(r,\theta,z)$ is defined, with $r=\sqrt{x^{2}+y^{2}}$ and
$\theta=\tan^{-1}y/x$.  The Green's function for the field at a point
$(x,y,z)$ due to a source of time dependence $\exp[-\J\omega t]$ at
$(x_{0},y_{0},z_{0})$ is given by~\cite{roger-moreau-kucukcoskun16}
\begin{align}
  \label{equ:analysis:greens:1}
  G(\mathbf{x}, \mathbf{x}_{0})
  &=
  \frac{\E^{-\J K M_{0}(X - X_{0})}}{\beta}
  \left[    
    g_{1}(\mathbf{x}, \mathbf{x}_{0}) +
    g_{2}(\mathbf{x}, \mathbf{x}_{0})
  \right],\\
  \label{equ:analysis:greens:2}
  g_{i}(\mathbf{x}, \mathbf{x}_{0}) &=
  \frac{\J K}{8\pi}
  \int_{-\infty}^{S_{i}}
  \frac{H_{1}^{(1)}(K \rbar_{i}\sqrt{1+u^{2}})}{\sqrt{1+u^{2}}}\,\D u,
\end{align}
where $H_{n}^{(1)}(\cdot)$ is the Hankel function of the first kind, and
\begin{align*}
  \beta &= \sqrt{1-M_{0}^{2}},\quad X = \frac{x}{\beta},\quad
  K = \frac{\omega}{c}\frac{1}{\beta},\\
  \rbar_{1,2} &=
  \left[
    (X-X_{0})^{2} + (y\mp y_{0})^{2} + (z-z_{0})^{2}
  \right]^{1/2},\\
  &=
  \left[
    \rbar^{2} + \rbar_{0}^{2} -
    2\rbar\rbar_{0}\cos
    \left(
      \tbar \mp \tbar_{0}
    \right)
    +
    (z-z_{0})^{2}
  \right]^{1/2},\\
  S_{1} &=
  \frac{2\sqrt{\rbar\rbar_{0}}}{\rbar_{1}}
  \cos\frac{\tbar - \tbar_{0}}{2},\quad
  S_{2} =
  -\frac{2\sqrt{\rbar\rbar_{0}}}{\rbar_{2}}
  \cos\frac{\tbar + \tbar_{0}}{2},\\
  \rbar &= \sqrt{X^{2}+y^{2}},\quad
  \tbar = \tan^{-1}\frac{y}{X}.
\end{align*}
For convenience, we note that the form
of~(\ref{equ:analysis:greens:1}) is that of calculating the sound from
a source and its image in the plane $y=0$. If we take $\theta_{1}$ as
the coordinate of the source and define $\theta_{2}=\theta_{1}-2\pi$
as the coordinate of the image, the acoustic field can be calculated
as the sum of the fields $g_{i}$ from the source and its image, using
\begin{align}
  \label{equ:analysis:reduced}
  \rbar_{i} &=
  \left[
    \rbar^{2} + \rbar_{0}^{2} -
    2\rbar\rbar_{0}\cos
    \left(
      \tbar - \tbar_{i}
    \right)
    +
    (z-z_{0})^{2}
  \right]^{1/2},\\  
  S_{i} &=
  \frac{2\sqrt{\rbar\rbar_{0}}}{\rbar_{i}}
  \cos\frac{\tbar - \tbar_{i}}{2}.
\end{align}

We rewrite~(\ref{equ:analysis:greens:2}) in a form which facilitates
physical interpretation when applied to the rotor noise problem, and
can be used to efficiently approximate the field in different
regions. Using the result
\begin{align}
  \int_{0}^{\infty}
  \frac{H_{1}^{(1)}(K \rbar_{i}\sqrt{1+u^{2}})}{\sqrt{1+u^{2}}}\,
  \D u
  &=
  -\J\frac{\E^{\J K\rbar_{i}}}{K\rbar_{i}},\nonumber
\end{align}
we can write $g_{i}$ in a number of alternative forms,
\begin{align}
  \label{equ:analysis:greens:3}
  g_{i}(\mathbf{x},\mathbf{x}_{0})
  &=
  \frac{\E^{\J K\rbar_{i}}}{4\pi\rbar_{i}}
  -
  \frac{\J K}{8\pi}\lfunc'(K\rbar_{i}, S_{i}),\\
  \label{equ:analysis:greens:4}
  &=
  \frac{\E^{\J K\rbar_{i}}}{8\pi\rbar_{i}}
  + \frac{\J K}{8\pi}\lfunc(K\rbar_{i}, S_{i}),    
\end{align}
where for convenience, we define the functions
\begin{align}
  \label{equ:analysis:lfunc}
  \lfunc(t, s)
  &=
  \int_{0}^{s}
  \frac{H_{1}^{(1)}(t\sqrt{1+u^{2}})}{\sqrt{1+u^{2}}}
  \,\D u,\\
  \label{equ:analysis:lfunc:d}
  \lfunc'(t, s)
  &=
  \int_{s}^{\infty}
  \frac{H_{1}^{(1)}(t\sqrt{1+u^{2}})}{\sqrt{1+u^{2}}}
  \,\D u.  
\end{align}

The evaluation of the Green's function for scattering by the plate
requires the evaluation of the integrals of~(\ref{equ:analysis:lfunc})
and~(\ref{equ:analysis:lfunc:d}). A solution for most of the far field
will be found by approximating~(\ref{equ:analysis:lfunc:d}), while an
exact expansion for~(\ref{equ:analysis:lfunc}) will be developed and
used to find a solution for the remaining part of the field.

Integration by parts gives an asymptotic approximation for
$\lfunc'(\cdot)$,
\begin{align}
  \label{equ:analysis:lfunc:a}
  \int_{s}^{\infty}
  \frac{H_{1}^{(1)}(t\sqrt{1+u^{2}})}{\sqrt{1+u^{2}}}\,\D u
  &\sim
  \frac{H_{0}^{(1)}(t\sqrt{1+s^{2}})}{t s},\quad t\to\infty,
\end{align}
so that for large $K\rbar_{i}$,
\begin{align}
  g_{i}(\mathbf{x},\mathbf{x}_{0})
  \label{equ:analysis:far:field}
  &\approx
  -\J\frac{H_{0}^{(1)}(Kv)}{8\pi|\rbar_{i}S_{i}|}
  +
  \frac{\E^{\J K \rbar_{i}}}{4\pi\rbar_{i}}H(S_{i}),\\
  v^{2} &= (\rbar+\rbar_{0})^{2} + (z-z_{0})^{2},\nonumber
\end{align}
and $H(\cdot)$ is the Heaviside step function. 

\begin{figure}[t]
  \includegraphics{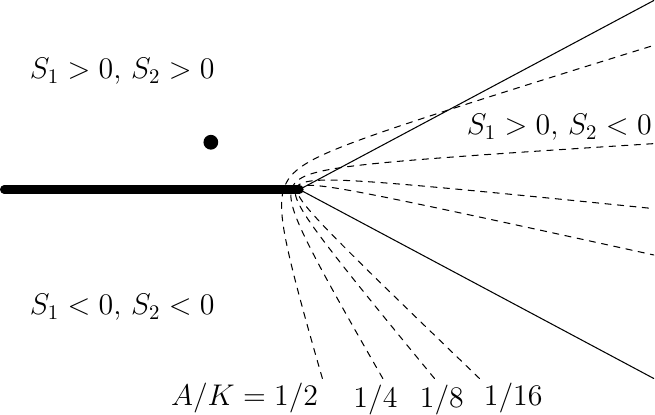}
  \caption{Contribution of acoustic source and image in different
    regions. Thick line: plate; solid lines: $S_{1,2}=0$; dashed lines
    hyperbolae of~(\ref{equ:analysis:hyper}).}
  \label{fig:analysis:regions}
\end{figure}

The asymptotic expression~(\ref{equ:analysis:far:field}) is clearly
unbounded for $|S_{i}|\to0$ and is invalid in the vicinity of
$S_{1,2}=0$. From the definition of $S_{i}$, it is clear that this is
the region around the lines through the source and the trailing edge
and through the image source and the edge,
Figure~\ref{fig:analysis:regions}. 

Jones~\cite{jones77} discusses the behaviour of the acoustic field in
different regions in terms of geometric acoustics. Where $S_{i}>0$,
the free-field Green's function is ``switched on'' and the field is
that of a source and its reflection in the plane. Where $S_{i}<0$, the
plate blocks the direct contribution from the source and only the
second term of~(\ref{equ:analysis:far:field}) contributes, defining
the ``shadow'' region. Near the boundaries of these two regions, the
asymptotic expression breaks down. In the notation of this paper, the
asymptotic expression is reasonable outside a region bounded by the
hyperbola~\cite{jones77}
\begin{align}
  \label{equ:analysis:hyper}
  K(\rbar + \rbar_{0} - \rbar_{1}) = A,
\end{align}
with Jones taking $A=10$ as a reasonable value. As well as the lines
$S_{i}=0$, Figure~\ref{fig:analysis:regions} shows the hyperbola
of~(\ref{equ:analysis:hyper}) for different values of $A/K$. The axis
of the hyperbolae is $S_{i}=0$ and they contract with increasing
$K$. At low frequency the asymptotic approximation breaks down in a
large region of the field. As $K$ increases, geometric acoustics is a
better approximation and more of the field is well modelled by the
asymptotic formula.

In order to find an expansion which can be used for small $S_{i}$, we
apply the product theorem for Bessel
functions~\cite[8.530.2]{gradshteyn-ryzhik80}
to~(\ref{equ:analysis:greens:4}),
\begin{align}
  \label{equ:analysis:series:1}  
  \frac{H_{1}^{(1)}(t\sqrt{1+u^{2}})}{\sqrt{1+u^{2}}}
  &=
  2\sum_{q=0}^{\infty}
  (-1)^{q}
  (2q+1)
  H_{2q+1}^{(1)}(t)
  \frac{J_{2q+1}(t u)}{t u}.
\end{align}
Upon integration of the Bessel functions of the first kind,
\begin{align}
  \label{equ:analysis:series:2}
  \lfunc(t, s)
  &= 
  \frac{2}{t}  
  \sum_{q=0}^{\infty}
  (-1)^{q}
  H_{2q+1}^{(1)}(t)
  \left[
    J_{2q+1}(t s)
    +
    2\sum_{k=1}^{\infty} J_{2q+2k+1}(t s)
  \right],
\end{align}
which is exact for $|s|<1$, though slow to converge as $|s|\to
1$. This result can be used to evaluate the acoustic field in the
region $s\to0$.

\subsection{Acoustic field of a ring source near a trailing edge}
\label{sec:analysis:ring}

\begin{figure}[t]
  \includegraphics{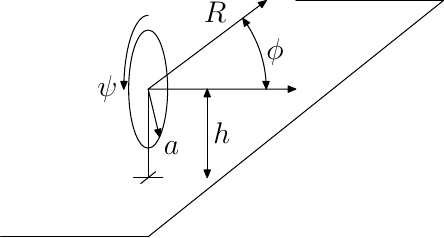}
  \caption{Coordinate system for evaluation of acoustic field of ring
    source}
  \label{fig:analysis:ring}
\end{figure}

We now apply the results of Section~\ref{sec:analysis:greens} to the
estimation of the acoustic field of a rotating source impinging on the
edge of a semi-infinite plate. The system is modelled as a ring source
and its image in the plane, with the field of each calculated using
the ``partial'' Green's function $G_{1,2}$ with
\begin{align}
  \label{equ:analysis:partial}
  G_{i}(\mathbf{x},\mathbf{x}_{0})
  &=
  \frac{\E^{-\J K M_{0}(X - X_{0})}}{\beta}g_{i}(\mathbf{x},\mathbf{x}_{0})
\end{align}
with the principal source denoted by subscript~$1$ and its image by
subscript~$2$.

As is standard for rotor noise
problems~\cite{gutin48,garrick-watkins53,wright69,carley06a,carley10},
the source is taken to vary as $\exp[\J n\psi_{0}]$ where $\psi_{0}$
is the angular variable on a circular source.  To respect the image
system, the image ring source has azimuthal variation $\exp[-\J
n\psi_{0}]$ and its field can be calculated using the same analysis,
changing $n$ to $-n$.

The ring is centred at a point $(x_{0}, h, 0)$ with coordinates on the
source given by
\begin{align*}
  \mathbf{x}_{0} &= (x_{0}, h + a\cos\psi_{0}, a\sin\psi_{0}),
\end{align*}
where $a$ is the radius of the source. We introduce spherical
coordinates centred on each ring source,
\begin{align*}
  X-X_{0} &= R_{i}\cos\phi_{i},\\
  y\mp h &= R_{i}\sin\phi_{i}\cos\psi_{i},\\
  z   &= R_{i}\sin\phi_{i}\sin\psi_{i}.
\end{align*}

The acoustic field of each source is then given by
\begin{align}
  \label{equ:analysis:4}
  p_{i}(\mathbf{x}) &=
  \frac{\E^{-\J K M_{0}(X - X_{0})}}{\beta}
  \int_{0}^{2\pi}
  \E^{\J n\psi_{0}}g_{i}(\mathbf{x}, \mathbf{x}_{0})
  \,\D\psi_{0}.
\end{align}
We now develop an approximation for $p_{i}$ valid in the far field.

We treat the field as made up of two contributions, corresponding to
the ``direct'' field generated by the ring source, and that scattered
by the trailing edge. The ``ring'' term is the familiar integral
\begin{align}
  \label{equ:rfunc}
  \rfunc_{n}(K, a, R_{i}, \phi_{i}, \psi_{i}) &= 
  \int_{0}^{2\pi}
  \frac{\E^{\J(K\rbar_{i}+n\psi_{0})}}{4\pi\rbar_{i}}\,\D\psi_{0},
\end{align}
for which a number of exact~\cite{carley10} and
approximate~\cite{gutin48,wright69} expressions exist.

To evaluate the part of the field scattered by the trailing edge, we
define the functions,
\begin{align}
  \label{equ:efunc:1}
  \efunc_{n}(K, a, \rbar_{0}, \rbar_{i}, S_{i})
  &=
  \frac{\J K}{8\pi}  
  \int_{0}^{2\pi}
  \E^{\J n\psi_{0}}
  \lfunc(K\rbar_{i}, S_{i})
  \D\psi_{0};
  \\
  \label{equ:efunc:2}
  \efunc'_{n}(K, a, \rbar_{0}, \rbar_{i}, S_{i})
  &=
  \frac{\J K}{8\pi}  
  \int_{0}^{2\pi}
  \E^{\J n\psi_{0}}
  \lfunc'(K\rbar_{i}, S_{i})\,
  \D\psi_{0}.
\end{align}

In the spherical coordinate system, $\rbar_{i}$ is given by
\begin{align*}
  \rbar_{i}^{2} &=
  R_{i}^{2} + a^{2} - 2a(y\mp h)\cos\psi_{0} - 2az\sin\psi_{0},\\
  &= R_{i}^{2} + a^{2} - 2R_{i}a\sin\phi_{i}\cos(\psi_{0}-\psi_{i}).
\end{align*}
Differentiating with respect to $a$ and neglecting terms less than
$O(1/R_{i})$ yields a far-field approximation,
\begin{align*}
  \rbar_{i} &\approx R_{i} - a\sin\phi_{i}\cos(\psi_{0}-\psi_{i}),\\
  \frac{1}{\rbar_{i}} &\approx \frac{1}{R_{i}},
\end{align*}
which, upon insertion into~(\ref{equ:analysis:4}) and use of
tables~\cite[8.411]{gradshteyn-ryzhik80}, yields the familiar result
for the acoustic field of a ring source~\cite[among many
others]{gutin48,garrick-watkins53,wright69},
\begin{align*}
  \rfunc_{n}(K, a, R_{i}, \phi_{i}, \psi_{i})
  &\approx
  (-\J)^{n}\E^{\J n\psi_{i}}\frac{\E^{\J K R_{i}}}{2 R_{i}}J_{n}(Ka\sin\phi_{i}),
\end{align*}
where $J_{n}(\cdot)$ is the Bessel function of the first kind.  The
free-field part of the ring-source field is then given by
\begin{align}
  \label{equ:ring:free:field}
  p_{i}^{(f)}(\mathbf{x}) &\approx
  \frac{\E^{-\J K M_{0}(X - X_{0})}}{\beta}\rfunc_{n}(K, a, R_{i},
  \phi_{i}, \psi_{i}).
\end{align}

We can apply a similar approach to the approximation of the scattered
field. Given
\begin{align*}
  v^{2} =
  \rbar_{1}^{2}(1+S_{1}^{2}) =
  \rbar_{2}^{2}(1+S_{2}^{2})
  &=
  \left(\rbar + \rbar_{0}\right)^{2} + (z-z_{0})^{2},
\end{align*}
to first order in $a$,
\begin{align*}
  v &\approx v_{0} + \frac{hSa}{v_{0}}\cos(\psi_{0}+\alpha),\\
  v_{0} &=
  \left[
    \left(
      \rbar + \sqrt{X_{0}^{2}+h^{2}}
    \right)^{2} + z^{2}
  \right]^{1/2},\\
  \alpha &= \tan^{-1}\frac{z/h}{1+\rbar/\sqrt{X_{0}^{2}+h^{2}}},\\
  S^{2} &=
  \left(\frac{\rbar}{\sqrt{X_{0}^{2}+h^{2}}} + 1\right)^{2}
  +
  \left(\frac{z}{h}\right)^{2}.
\end{align*}

Using Neumann's addition theorem~\cite[10.23]{dlmf10},
\begin{align*}
  H_{0}^{(1)}(Kv) &\approx
  H_{0}^{(1)}\left(Kv_{0} + \frac{KhSa}{v_{0}}\cos(\psi_{0}+\alpha)\right),\\
  &=
  \sum_{q=-\infty}^{\infty} H_{q}^{(1)}(K v_{0})
  J_{-q}\left(K\frac{hSa}{v_{0}}\cos(\psi_{0}+\alpha)\right).
\end{align*}
With the use of tables~\cite[6.681.8,~9]{gradshteyn-ryzhik80}, it is
readily shown that for $n$ and $q$ both odd or both even, the integral
\begin{align}
  \label{equ:analysis:bessel:psi}
  \int_{0}^{2\pi}
  \E^{\J n\psi}
  J_{q}(\alpha\cos\psi)
  \,\D\psi
  &=
  2\pi J_{(q-n)/2}(\alpha/2)J_{(q+n)/2}(\alpha/2),
\end{align}
and is identically zero otherwise. This yields
\begin{align*}    
  \int_{0}^{2\pi} H_{0}^{(1)}(Kv) \E^{\J n\psi_{0}}\,\D\psi_{0}
  &\approx
  2\pi\E^{-\J n\alpha}
  \sum_{q=-\infty}^{\infty}
  H_{2q+\delta}^{(1)}(K v_{0})
  J_{-(n'+q+\delta)}\left(K\frac{hSa}{2v_{0}}\right)
  J_{(n'-q)}\left(K\frac{hSa}{2v_{0}}\right),\\  
\end{align*}
where $n=2n'+\delta$, $\delta=0,1$.

The part of the ring-source field scattered by the trailing edge,
\begin{align*}
  \efunc'_{n}(K, a, \rbar_{0}, \rbar_{i}, S_{i})
  &=
  \frac{\J K}{8\pi}  
  \int_{0}^{2\pi}
  \E^{\J n\psi_{0}}
  \lfunc'(K\rbar_{i},S_{i})\,
  \D\psi_{0},
\end{align*}
is then approximated by
\begin{align}
  \label{equ:ring:scattered}
  \efunc'_{n}(K, a, \rbar_{0}, \rbar_{i}, S_{i})
  &\approx
  -\frac{\J}{4}\frac{\E^{-\J n\alpha}}{\sqrt{v_{0}^{2}-R_{i}^{2}}}
  \sum_{q=-\infty}^{\infty}
  H_{2q+\delta}^{(1)}(K v_{0})
  J_{-(n'+q+\delta)}\left(K\frac{hSa}{2v_{0}}\right)
  J_{(n'-q)}\left(K\frac{hSa}{2v_{0}}\right).
\end{align}
This approximation breaks down between the free-field being switched
on or off where $S_{i}$ is small ($v_{0}\approx R_{i}$) and a
different approximation is required in this region.

Again, we proceed by approximating for the far field and using the
properties of Bessel functions to find a form suitable for evaluation.

Returning to~(\ref{equ:analysis:series:2}),
\begin{align*}
  \lfunc(\alpha, s)
  &= 
  \frac{2}{\alpha}  
  \sum_{q=0}^{\infty}
  (-1)^{q}
  H_{2q+1}^{(1)}(\alpha)
  \left[
    J_{2q+1}(\alpha s)
    +
    2\sum_{k=1}^{\infty} J_{2q+2k+1}(\alpha s)
  \right],
\end{align*}
and noting that $\alpha=K\rbar_{i}$, we seek an approximation which
can be inserted into~(\ref{equ:analysis:4}) and integrated. In the far
field, $K\rbar_{i}\to\infty$, the difficulty lies in capturing the
rapid variation in $K\rbar_{i}s_{i}$ as $s_{i}\to0$. We proceed as
follows to develop a tractable far-field approximation. The Hankel
function is replaced by an approximation valid to first order in $a$,
using the standard relations for derivatives,
\begin{align*}
  H_{m}(K\rbar_{i})
  &\approx
  H_{m}(KR_{i})
  + Ka H_{m+1}(KR_{i})\sin\phi_{i}\cos(\psi_{0}-\psi_{i})
  + O(1/(KR_{i})^{2}).
\end{align*}
The argument of the Bessel function of the first kind is approximated
by
\begin{align*}
  K\rbar_{i}s_{i}
  &\approx
  2K\sqrt{\rbar\rbar_{0}}\cos\frac{\tbar-\tbar_{0}}{2}
  +
  Ka \sqrt{\frac{\rbar}{\rbar_{0}}}
  \sin\frac{\tbar+\tbar_{0}}{2}\cos\psi_{0},
\end{align*}
and again using the product
theorem~\cite[8.530.2]{gradshteyn-ryzhik80},
\begin{align*}
  J_{m}(K\rbar_{i}s_{i})
  &\approx
  \sum_{k=-\infty}^{\infty}
  J_{m-k}
  \left(
    2K\sqrt{\rbar\rbar_{0}}\cos\frac{\tbar-\tbar_{0}}{2}
  \right)
  J_{k}
  \left(
    Ka\sqrt{\frac{\rbar}{\rbar_{0}}}\sin\frac{\tbar+\tbar_{0}}{2}
    \cos\psi_{0}
  \right),
\end{align*}
yielding
\begin{align}
  \label{equ:analysis:edge:lfunc}
  \lfunc(K\rbar_{i},s_{i})
  &\approx
  \frac{2}{KR_{i}}
  \sum_{q=0}^{\infty}
  (-1)^{q}
  \left[
    H_{2q+1}^{(1)}(KR_{i})
    +Ka\sin\phi_{i}H_{2q+2}^{(1)}(KR_{i})\cos(\psi_{0}-\psi_{i})
  \right]\nonumber\\
  &\times
  \biggl\{
    \sum_{m=-\infty}^{\infty}
    J_{m}(\gamma_{2}\cos\psi_{0})
    \left[
      J_{2q+1-m}\left(\gamma_{1}\right)
      +2\sum_{k=1}^{\infty}
      J_{2q+2k+1-m}(\gamma_{1})
    \right]
    \biggr\},\\
    \gamma_{1} &=
    2K\sqrt{\rbar\rbar_{0}}\cos\frac{\tbar-\tbar_{0}}{2},\quad
    \gamma_{2} =
    Ka\sqrt{\frac{\rbar}{\rbar_{0}}}\sin\frac{\tbar+\tbar_{0}}{2},\nonumber
\end{align}
where all source coordinates such as $\tbar$ are evaluated at $a=0$,
the centre of the ring. Using~(\ref{equ:analysis:bessel:psi}), for
$n=2n'$,
\begin{align}
  \efunc_{n}(K, a, \rbar_{0}, \rbar_{i}, S_{i})
  &=
  \frac{\J K}{8\pi}  
  \int_{0}^{2\pi}
  \lfunc(K\rbar_{i},s_{i})
  \E^{\J n\psi_{0}}\,\D\psi_{0}
  \approx\nonumber\\
  \frac{\J}{2R_{i}}
  \biggl\{
  &\sum_{q=0}^{\infty}
  (-1)^{q}H_{2q+1}^{(1)}(K R_{i})
  \sum_{m=-\infty}^{\infty}
  J_{m-n'}\left(\frac{\gamma_{2}}{2}\right)
  J_{m+n'}\left(\frac{\gamma_{2}}{2}\right)
  M_{1}(\gamma_{1})\nonumber\\
  +\E^{-\J\psi_{i}}
  \frac{Ka\sin\phi_{i}}{2}
  &\sum_{q=0}^{\infty}
  (-1)^{q}H_{2q+2}^{(1)}(K R_{i})
  \sum_{m=-\infty}^{\infty}
  J_{m-n'}\left(\frac{\gamma_{2}}{2}\right)
  J_{m+n'+1}\left(\frac{\gamma_{2}}{2}\right)
  M_{2}(\gamma_{1})\nonumber\\
  +
  \E^{\J\psi_{i}}
  \frac{Ka\sin\phi_{i}}{2}
  &\sum_{q=0}^{\infty}
  (-1)^{q}H_{2q+2}^{(1)}(K R_{i})
  \sum_{m=-\infty}^{\infty}
  J_{m-n'}\left(\frac{\gamma_{2}}{2}\right)
  J_{m+n'-1}\left(\frac{\gamma_{2}}{2}\right)
  M_{3}(\gamma_{1})
  \biggr\}
  \label{equ:analysis:edge:func},    
\end{align}
where the auxiliary quantities are
\begin{align*}
  &M_{1}(\gamma_{1})
  =
  J_{2(q-m)+1}(\gamma_{1}) +
  2\sum_{k=1}^{\infty}J_{2(q+k-m)+1}(\gamma_{1}),\nonumber\\
  &M_{2}(\gamma_{1})
  =
  J_{2(q-m)}(\gamma_{1}) +
  2\sum_{k=1}^{\infty}J_{2(q+k-m)}(\gamma_{1}),\nonumber\\
  &M_{3}(\gamma_{1})
  =
  J_{2(q-m+1)}(\gamma_{1}) +
  2\sum_{k=1}^{\infty}J_{2(q+k-m+1)}(\gamma_{1}).
\end{align*}

\subsection{Summary}
\label{sec:analysis:summary}

\begin{figure}
  \centering
  \includegraphics{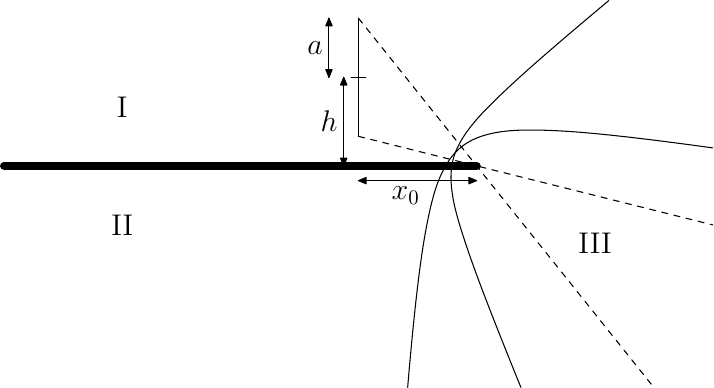}
  \caption{Regions around rotor labelled by far-field
    approximation. The hyperbolae of~(\ref{equ:analysis:hyper}) are
    shown for the top and bottom of the rotor with $A=10$ and
    $K=12/\sqrt{1-0.15^2}$.}
  \label{fig:summary:regions}
\end{figure}

To summarize, the far-field sound from a rotor near the edge of a
semi-infinite plane can be estimated by summing the contribution from
the circular source of azimuthal order $n$ and from its image in the
plane, of order $-n$. For each source, the acoustic field is estimated
by first determining the appropriate far-field approximation to
apply. Figure~\ref{fig:summary:regions} shows the region around a
rotor divided according to the sign of $S_{i}$, with the hyperbolae
of~(\ref{equ:analysis:hyper}) added for source positions at the top
and bottom of the rotor, with $A=10$ and $K=12/\sqrt{1-0.15^2}$, which
corresponds to the sample calculations presented in
Section~\ref{sec:results}. In region~I, $S_{i}>0$ for all source
positions on the rotor while in region~II, $S_{i}<0$. In region~III,
$S_{i}$ changes sign for some value of $\psi_{0}$. The appropriate
far-field approximation to use in each region is then:
\begin{subequations}
  \begin{alignat}{2}
    \label{equ:summary:1}
    p_{n} &= \frac{\E^{-\J K M_{0}(X - X_{0})}}{\beta}
    \left(\rfunc_{n} - \efunc_{n}'\right), & \text{(I)}\\
    \label{equ:summary:2}
    &= -\frac{\E^{-\J K M_{0}(X - X_{0})}}{\beta}
    \efunc_{n}', & \text{(II)}\\
    &= \frac{\E^{-\J K M_{0}(X - X_{0})}}{\beta}
    \left(\frac{\rfunc_{n}}{2} + \efunc_{n}\right),\quad &
    \text{(III)}.
    \label{equ:summary:3}
  \end{alignat}
\end{subequations}
In the far field, the functions $\rfunc(\cdot)$, $\efunc(\cdot)$, and
$\efunc'(\cdot)$ can be evaluated using the results given
earlier. Finally, with regard to evaluation of the series defining
these functions, we note that the Bessel function of the first kind
$J_{n}(x)$ decays exponentially with increasing order for $n>x$,
allowing the series to be truncated with negligible loss of accuracy. 

\section{Results}
\label{sec:results}

To illustrate the application of the method, we present sample results
for an arbitrarily selected configuration, with rotor parameters
$a=0.3$, $n=6$. The rotor is positioned above the plate at $h=0.9$,
and one rotor diameter upstream of the trailing edge at
$x_{0}=-2a$. The rotor tip Mach number $M_{t}$ is set to~0.6, so that
the wavenumber $k=n M_{t}/a=12$, and the flow Mach number
$M_{0}=0.15$. The acoustic field is evaluated using the methods of
this paper, and compared to full numerical evaluation, with results
calculated above and below the plate to examine the shielding effect.

\begin{figure}
  \centering
  \includegraphics{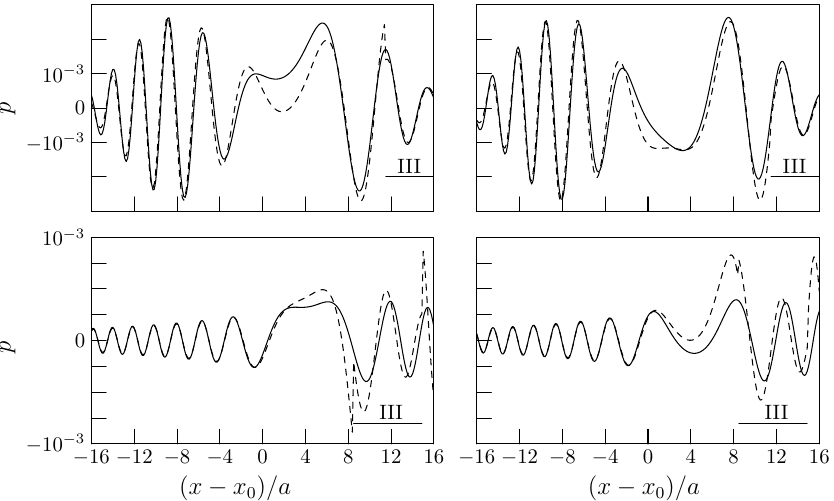}
  \caption{Acoustic potential above and below source; left-hand
    column: real part; right-hand column: imaginary part; solid line:
    numerical; dashed line: approximation. Top row: field above source
    and plate; bottom row: field below source and plate.}
  \label{fig:results:potential}
\end{figure}

Figure~\ref{fig:results:potential} shows the real and imaginary part
of the potential at $-16a\leq x-x_{0}\leq 16a$, $z=0$, i.e.\ directly
above and below the source centre. Results are calculated at a
constant vertical displacement above and below the source,
$y=h\pm 16a$. The two upper plots show the acoustic potential above
the source and the plate, and the lower plots show the field below the
plate, at the same distance from the source as in the upper plots.
The region marked ``III'' on each plot shows where
(\ref{equ:summary:3}) has been applied, in the transition region where
the asymptotic approximation for the edge-scattered field breaks
down. The beginning of this breakdown is apparent at the end of
region~III, in particular in the lower left-hand plot of the figure,
where the difference between the asymptotic approximation and the
numerical evaluation is clear for $(x-x_{0})/a\gtrapprox0$. This
corresponds to the hyperbolic boundary shown in
Figure~\ref{fig:summary:regions}, which has been drawn for the same
value of $K$ as used in the calculations. 

In applications, where the rotor may be placed above a wing in order
to benefit from shielding effects, the region of interest is the
``shadow zone'' upstream of the trailing edge. The upper and lower
fields are computed at constant distance from the source, so the
different scales on the upper and lower plots make the shielding
effect clear. In this case, the maximum amplitude upstream of the
trailing edge is about ten times greater above the plate than below,
and is well predicted by the far field approximation. 

We note that while the results presented demonstrate good accuracy,
the analysis is limited by the assumptions made in its development. In
particular, the analysis does assume small source radius $a$ and
numerical experimentation has found that the accuracy of predictions
degrades as $a$ is increased. This may limit the applicability of the
method, in particular as $a$ becomes comparable to $h$, the case of a
rotor with a small tip clearance above the wing, or where the rotor
lies close to the trailing edge.

\section{Conclusions}
\label{sec:conclusions}

An approximate analysis has been developed for the radiation of sound
from a rotating source near the edge of a semi-infinite plate in a
uniform flow, a model problem for shielding and scattering of rotor
noise by a wing. Sample calculations for small source radius show good
agreement with full numerical evaluation, and capture the wing
shielding effect reasonably accurately. Future work will focus on
improving the accuracy and range of validity of the approximation, in
particular in the case of small clearance between the rotor and
wing. An open question is that of whether there is a useful exact
expansion for the acoustic field, comparable to the series solutions
which exist for isolated rotors.

\section*{Acknowledgements}

The author is indebted to the anonymous reviewer who brought the paper
of Jones~\cite{jones77} to his attention. 


\end{document}